\documentclass[letterpaper]{article} 
\usepackage{aaai24}  
\usepackage{times}  
\usepackage{helvet}  
\usepackage{courier}  
\usepackage[hyphens]{url}  
\usepackage{graphicx} 
\usepackage{natbib}  
\usepackage{caption} 
\usepackage{algorithm}
\usepackage{algorithmic}

\usepackage{subcaption}
\usepackage{bm}
\usepackage{multirow}
\usepackage{amsfonts}
\usepackage{amsmath}

\usepackage{newfloat}
\usepackage{listings}
\DeclareCaptionStyle{ruled}{labelfont=normalfont,labelsep=colon,strut=off} 
\floatstyle{ruled}
\newfloat{listing}{tb}{lst}{}
\floatname{listing}{Listing}
\title{CARAT: Contrastive Feature Reconstruction and Aggregation for Multi-Modal Multi-Label Emotion Recognition}
\author{
    Cheng Peng,
    Ke Chen \thanks{Ke Chen and Gang Chen are the corresponding authors.},
    Lidan Shou,
    Gang Chen ${}^*$\
}
\affiliations{
    The State Key Laboratory of Blockchain and Data Security, Zhejiang University, Hangzhou 310000, China \\
    \{chengchng, chenk, should, cg\}@zju.edu.cn
}

\usepackage{bibentry}

\begin{document}

\maketitle

\begin{abstract}
Multi-modal multi-label emotion recognition (MMER) aims to identify relevant emotions from multiple modalities. The challenge of MMER is how to effectively capture discriminative features for multiple labels from heterogeneous data. Recent studies are mainly devoted to exploring various fusion strategies to integrate multi-modal information into a unified representation for all labels. However, such a learning scheme not only overlooks the specificity of each modality but also fails to capture individual discriminative features for different labels. Moreover, dependencies of labels and modalities cannot be effectively modeled. To address these issues, this paper presents ContrAstive feature Reconstruction and AggregaTion (CARAT) for the MMER task. Specifically, we devise a reconstruction-based fusion mechanism to better model fine-grained modality-to-label dependencies by contrastively learning modal-separated and label-specific features. To further exploit the modality complementarity, we introduce a shuffle-based aggregation strategy to enrich co-occurrence collaboration among labels. Experiments on two benchmark datasets CMU-MOSEI and M3ED demonstrate the effectiveness of CARAT over state-of-the-art methods. Code is available at https://github.com/chengzju/CARAT.
\end{abstract}

\section{Introduction}

Multi-modal Multi-label Emotion Recognition (MMER) aims to identify multiple emotions (e.g., happiness and sadness) from multiple heterogeneous modalities (e.g., text, visual, and audio).
Over the last decades, MMER has fueled research in many communities, such as online chatting \cite{galik2012modelling}, news analysis \cite{zhu2019adversarial} and 
dialogue systems \cite{ghosal2019dialoguegcn}.

Different from single-modal tasks, multi-modal learning synergistically processes heterogeneous information from various sources, which introduces a challenge of how to capture discriminative representations from multiple modalities. 
To this end, recent works propose various advanced multi-modal fusion strategies to bridge the modality gap and learn effective representations \cite{ramachandram2017deep}.
According to the fusion manner, methods can be roughly divided into three categories: aggregation-based, alignment-based, and the mixture of them\cite{baltruvsaitis2018multimodal}.
The aggregation-based fusion employs averaging \cite{hazirbas2017fusenet}, concatenation \cite{ngiam2011multimodal} or attention \cite{zadeh2018multi} to integrate multi-modal features.
The alignment-based fusion \cite{pham2018seq2seq2sentiment, pham2019found} adopts the cross-modal adaptation to align latent information of different modalities.
However, unifying multiple modalities into one identical representation can inevitably neglect the specificity of each modality, thus losing the rich discriminative features.
Although recent works \cite{hazarika2020misa, zhang2022tailor} attempt to learn modality-specific representations, they still utilize attention to fuse these representations into one.
Therefore, a key challenge of MMER is how to effectively represent multi-modal data while maintaining modality specificity and integrating complementary information.

\begin{figure}[t]
\centering

\begin{subfigure}{0.65\linewidth}
    \centering
    \includegraphics[width=1.0\linewidth]{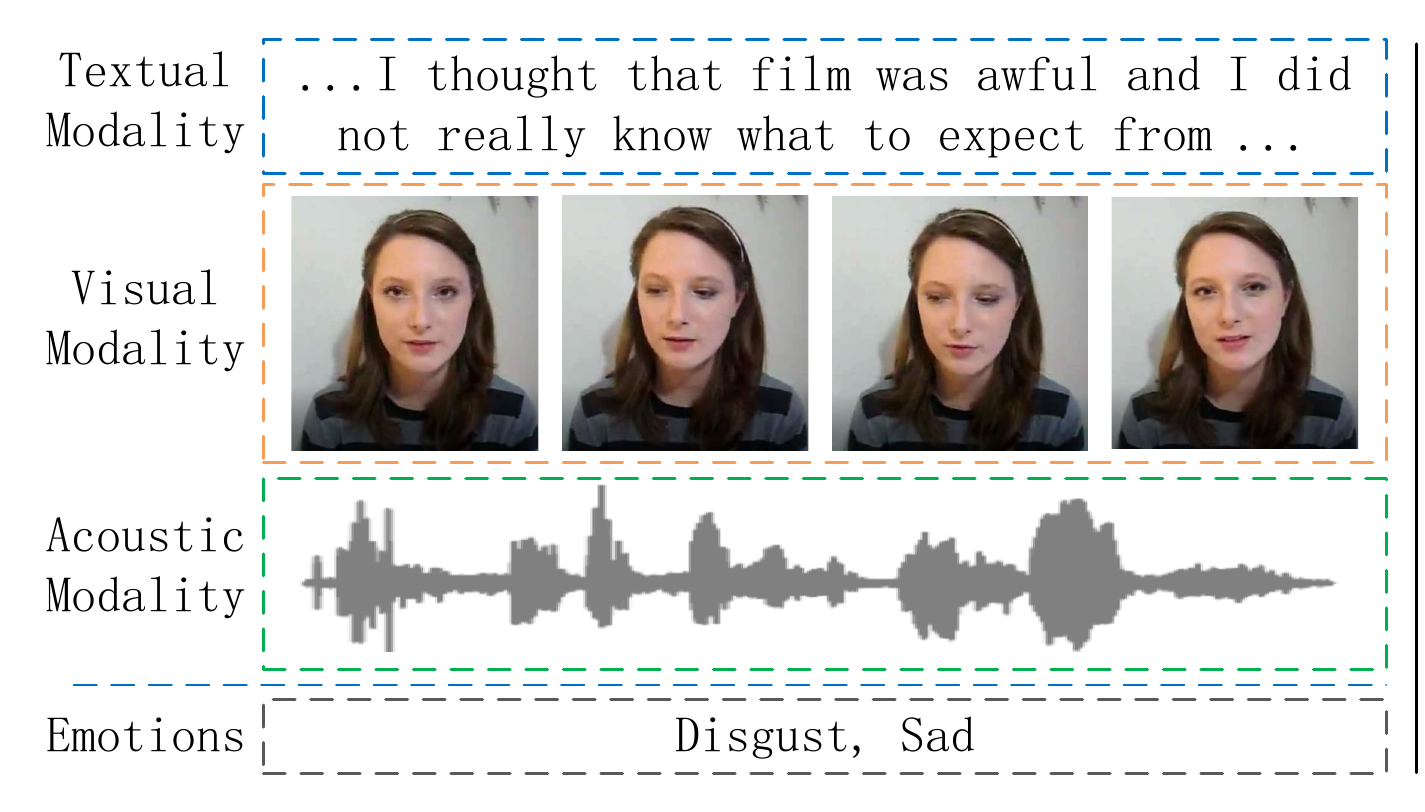}
\end{subfigure}
\begin{subfigure}{0.34\linewidth}
    \centering
    \begin{subfigure}{0.99\linewidth}
    \centering
    \includegraphics[width=0.99\linewidth]{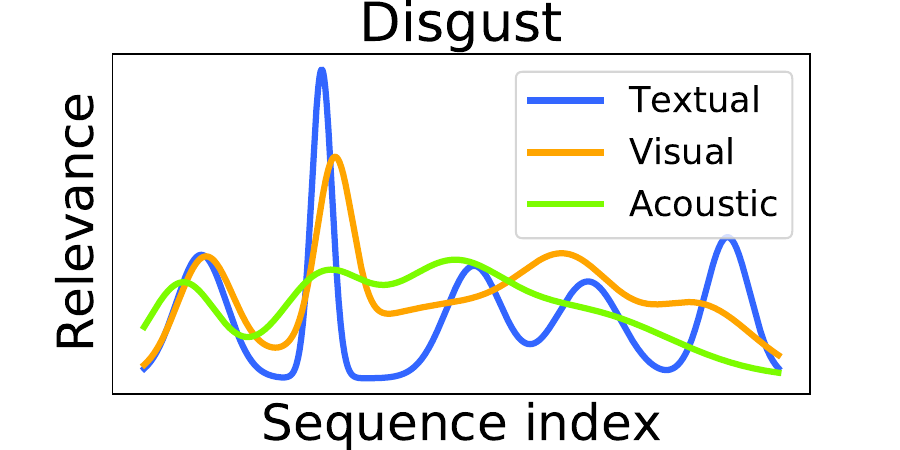}
\end{subfigure}
\begin{subfigure}{0.99\linewidth}
    \centering
    \includegraphics[width=0.99\linewidth]{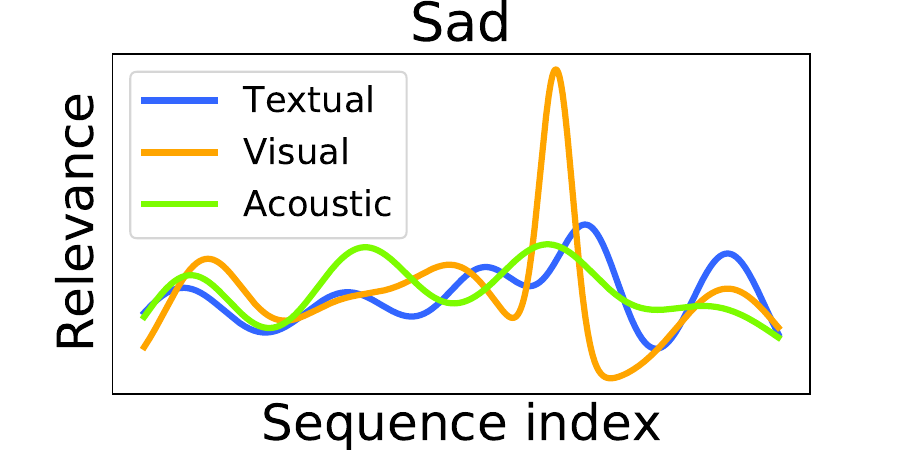}
\end{subfigure}
\end{subfigure}
\caption{ 
An example of MMER (left) and correlations between two relevant emotions and the video sequence (right).
}
\label{fig:intro_case}
\end{figure}

As a multi-label task \cite{zhang2013review}, MMER also needs to deal with complex dependencies among labels.
Nowadays, massive studies attempt various methods to explore label correlation, such as label similarity \cite{xiao2019label} and co-occurrence label graph \cite{ma2021label}.
However, these static correlations cannot reflect the collaborative relationship among labels.
On the other hand, another tricky conundrum for MMER is how to learn dependencies between labels and modalities.
Commonly, different modalities have inconsistent emotional expressions, and conversely, different emotions focus on different modalities, which means that inferring each potential label largely depends on the different contributions of different modalities.
As shown in Figure \ref{fig:intro_case}, we can infer \emph{sadness} more easily from the visual modality, while \emph{disgust} can be predicted from both textual and visual modalities.
Therefore, another challenge of MMER is how to effectively model both label-to-label and modality-to-label dependencies.

To address these issues, we propose ContrAstive feature Reconstruction and AggregaTion for MMER (CARAT), which coordinates representation learning and dependency modeling in a coherent and synergistic framework.
Specifically, our framework CARAT encapsulates three key components.
First, we adopt the label-wise attention mechanism to extract label-specific representations within each modality severally, which is intended to capture relevant discriminative features of each label while maintaining modality specificity.
Second, to reconcile the complementarity and specificity of multi-modal information, we develop an ingenious reconstruction-based fusion strategy that attempts to generate features of any modality by exploiting the information from multiple modalities. 
We leverage contrastive learning \cite{khosla2020supervised}, which is unexplored in previous MMER literature, to facilitate the learning of modal-separated and label-specific features.
Third, based on the reconstructed embeddings, we propose a novel sample-wise and modality-wise shuffle strategy to enrich the co-occurrence dependencies among labels.
After shuffled, embeddings are aggregated to finetune a robust discriminator.
Moreover, as for modeling the modality-to-label dependency, we employ a max pooling-like network to discover the most relevant modalities for different emotions per sample, and then impel these corresponding representations to be more discriminative. \footnote{This paper's complete version with technical appendices is available at \url{https://arxiv.org/abs/2312.10201}}
The main contributions of this paper can be summarized as follows:

\begin{itemize}
\item A novel framework, ContrAstive feature Reconstruction and AggregaTion, is proposed.
To the best of our knowledge, this work pioneers the exploitation of contrastive learning to facilitate a multi-modal fusion mechanism based on feature reconstruction.
As an integral part of our method, we also introduce a shuffle-based feature aggregation strategy, which uses the reconstructed embeddings to better leverage multi-modal complementarity.

\item To preserve the modality specificity, CARAT independently extracts label-specific representations from different modalities via label-wise attention. Then a max pooling-like network is involved to select the most relevant modal representation per emotion to explore potential dependencies between modalities and labels.

\item We conduct experiments on two benchmark datasets CMU-MOSEI and M${}^{3}$ED.
The experimental results demonstrate that our proposed method outperforms previous methods and achieves state-of-the-art performance.

\end{itemize}

\section{Related Works}
\begin{figure*}[t]
\centering
\includegraphics[width=0.999 \linewidth]{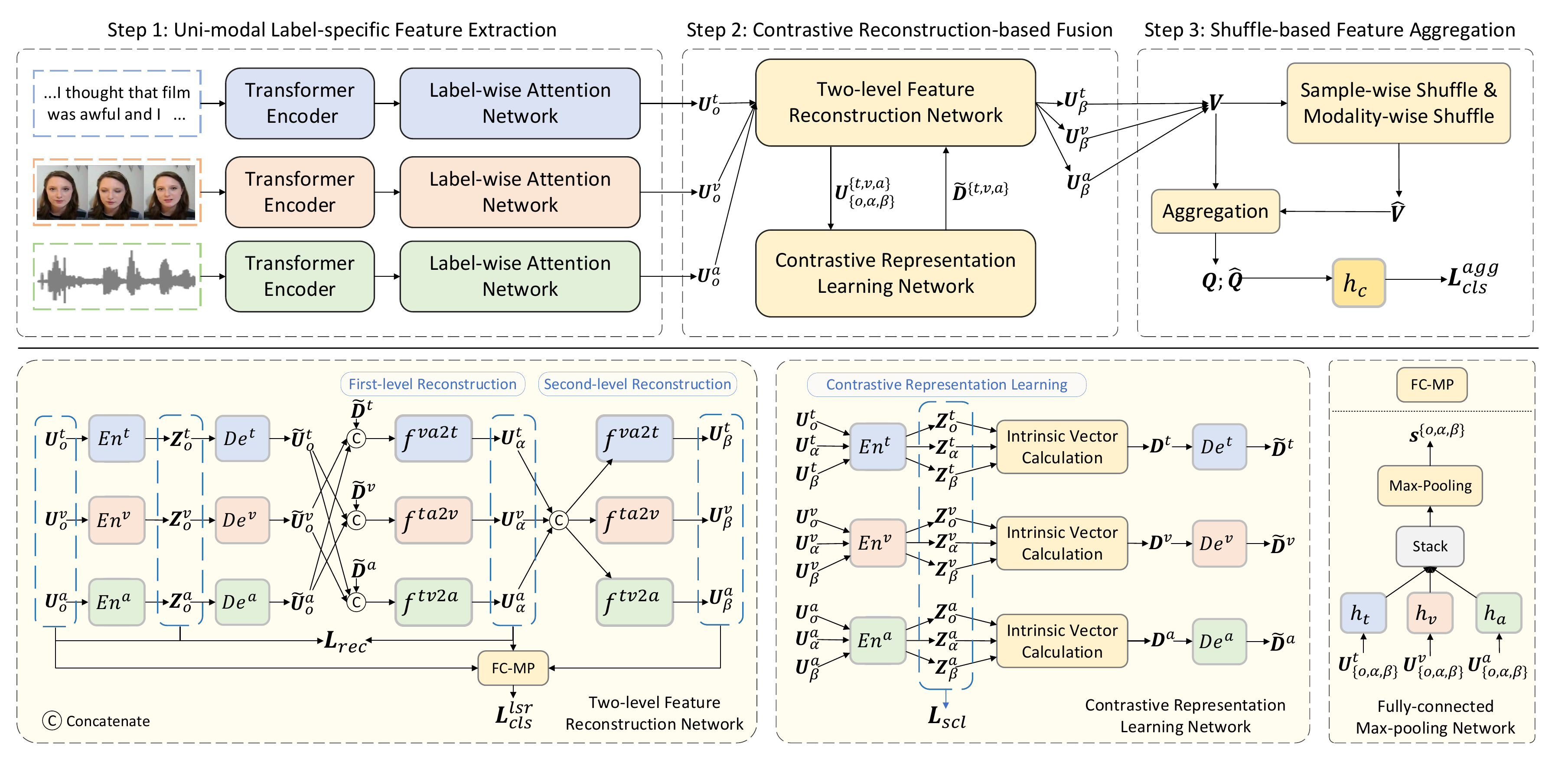} 
\caption{
The overall structure of CARAT with three sequential steps (up). 
Detailed implementations of two-level feature reconstruction network, contrastive representation learning network, and Max Pooling-like network (bottom).
}
\label{fig:method_overall}
\end{figure*}

\noindent{\textbf{Multi-modal Learning}} aims to build models that can process and relate information from multiple modalities \cite{baltruvsaitis2018multimodal}.
A fundamental challenge is how to effectively fuse multi-modal information.
According to the fusion manner, methods can be roughly divided into three categories: aggregation-based, alignment-based, and hybrid methods.
Aggregation-based methods use concatenation \cite{ngiam2011multimodal}, tensor fusion \cite{zadeh2017tensor, liu2018efficient} and attention \cite{zadeh2018multi} to combine multiple modalities, but suffer from the modality gap.
To bridge the gap, alignment-based fusion \cite{pham2018seq2seq2sentiment, pham2019found} exploits latent cross-modal adaptation by constructing a joint embedding space.
However, alignment-based fusion neglects the specificity of each modality,
resulting in the omission of discriminative information.

\noindent{\textbf{Multi-label Emotion Recognition}} is a foundational multi-label (ML) task and ML approaches can be quickly applied.
BR \cite{boutell2004learning} decomposes the ML task into multiple binary classification ones while ignoring label correlations.
To exploit the correlations, LP \cite{tsoumakas2006multi}, CC \cite{read2011classifier} and Seq2Seq \cite{yang2018sgm} are proposed.
To further explore label relationships, recent works leverage reinforced approach \cite{yang2019deep}, multi-task pattern \cite{tsai2020order}, and GCN model \cite{chen2019multi}.
Another important task is to learn effective label representations.
To compensate for the inability of a single representation to capture discriminative information of all labels, recent works \cite{chen2019learning, chen2019multi} utilize label-specific representations to capture the most relevant features for each label, which has been successfully applied to many studies \cite{huang2016learning, xiao2019label}.

\noindent{\textbf{Contrastive learning}} (CL) is an effective self-supervised learning technique 
\cite{li2020prototypical,oord2018representation,hjelm2018learning} .
CL aims to learn a discriminative latent space where similar samples are pulled together and dissimilar samples are pushed apart.
Motivated by the successful application of CL in unsupervised learning \cite{oord2018representation,he2020momentum},
Supervised Contrastive Learning (SCL) \cite{khosla2020supervised} is devised to promote a series of supervised tasks.
Recently, CL has been applied to multi-modal tasks to strengthen the interaction between features of different modalities \cite{zheng2022promptlearner, franceschini2022multimodal, zolfaghari2021crossclr}.
However, there has been no exploration of contrastive learning on multi-modal tasks in the multi-label scenario.

\section{Methodology}
In this section, we describe our CARAT framework, which comprises three sequential components (in Figure \ref{fig:method_overall}).

\subsection{Problem Definition}
We define notations for MMER.
Let $\mathcal{X}^{t} \!\in\! \mathbb{R}^{n_t \times d_t}$, $\mathcal{X}^{v} \!\in\! \mathbb{R}^{n_v \times d_v}$ and $\mathcal{X}^{a} \!\in \!\mathbb{R}^{n_a \times d_a}$ be the heterogeneous feature spaces for textual ($t$), visual ($v$) and acoustic ($a$) modality respectively, where $n_{m}$ and $d_{m}$ denotes the sequence length and modality dimension respectively ($m \in \{t,v,a\}$ is used to represent any modality).
And $\mathcal{Y}$ is the label space with $\mathcal{C}$ labels.
Given a training dataset $\mathcal{D} \!=\!\{(\bm{X}^{\{t,v,a\}}_i,\bm{y}_i)\}_{i=1}^{N}$,
MMER aims to learn a function $\mathcal{F}:\mathcal{X}^{t} \times \mathcal{X}^{v} \times \mathcal{X}^{a} \mapsto \mathcal{Y}$ to predict relevant emotions for each video.
Concretely, $\bm{X}_{i}^{m} \in \mathcal{X}^{m}$
are asynchronous coordinated utterance sequences and $\bm{y}_i\!=\! {\{0,1\}}^{\mathcal{C}}$ is the multi-hot label vector,
where sign $\bm{y}_{i,j}\!=\!1$ indicates that sample $i$ belongs to class $j$, otherwise $\bm{y}_{i,j}\!=\!0$.

\subsection{Uni-modal Label-specific Feature Extraction}

As the first step, this component aims to extract the relevant discriminative features for each label in each modality.

\subsubsection{\textbf{Transformer-based Extractor.}}
For each modality $m$, we use an independent Transformer Encoder \cite{vaswani2017attention} to map raw feature sequences $\bm{X}^m \!\in\! \mathbb{R}^{n_m \times d_m}$ into high-level embedding sequences $\bm{H}^m \!\in \!\mathbb{R}^{n_m \times d}$.
Each encoder is composed of $l_m$ identical layers, where each layer consists of two sub-layers: a multi-head self-attention sub-layer and a position-wise feed-forward sub-layer.
The residual connection \cite{he2016deep} is employed around each of the two sub-layers, followed by layer normalization.

\subsubsection{\textbf{Multi-label Attention.}}
Considering that each emotion is usually expressed by the most relevant part of the utterance,
we generate label-specific representations for each emotion to capture the most critical information.
After obtaining embedding sequences $\bm{H}^{m}$, 
we compute the combination of these embeddings for each label $j$ under each modality $m$ through a label-wise attention network.
Formally, we represent the hidden state of each embedding as $\bm{h}^m_{i}\! \in\! \mathbb{R}^{d} (i \!\in\! [n_m])$.
The attentional representation $\bm{u}_j^m$ is obtained as:
\begin{equation}
\bm{u}_{j}^m=\sum_{i=1}^{n_m} \alpha^{m}_{i j} \bm{h}_{i}^m, \quad \alpha^{m}_{i j}=\frac{\exp({{\bm{w}^{m}_{j}}^{\top} \bm{h}_{i}^m})}{\sum_{i^{\prime}=1}^{n_m} \exp({{\bm{w}^{m}_{j}}^{\top} \bm{h}^m_{i^{\prime}}})},
\end{equation}
where $\bm{w}^{m}_{j} \in \mathbb{R}^d$ denotes the attention parameter for the $j$-th label and $\alpha^{m}_{i j}$ is the normalized coefficient of $\bm{h}_{i}^m$.
It is worth noting that attention networks between modalities are still independent of each other, thus generating label-specific representations $\bm{U}^{m}_{o} \in \mathbb{R}^{\mathcal{C} \times d}$ separately.

\subsection{Contrastive Reconstruction-based Fusion}
The second component aims to utilize information from multiple modalities to restore the features of any modality.

\subsubsection{\textbf{Multi-modal Feature Reconstruction.}}
Considering that fusing multi-modal information into an identical representation can ignore the modality specificity, we propose a reconstruction-based fusion mechanism, which restores features of any modality with the feature distribution in the current modality and the semantic information in other modalities.
We first use three modality-specific encoders $En^{m}(\cdot)$ to project $\bm{U}^{m}_{o}$ into latent vectors $\bm{Z}_{o}^{m} \in \mathbb{R}^{\mathcal{C} \times d_z}$ in the latent space $\mathcal{S}^{z}$.
From the space $\mathcal{S}^{z}$, we calculate the intrinsic vectors $\bm{D}^{m} = \{\bm{d}_j^m \in \mathbb{R}^{d_z} \}_{j=1}^{\mathcal{C}}$ to reflect the feature distribution of each label $j$ in different modalities (explained in the next sub-section).
Then, three modality-specific decoders $De^{m}(\cdot)$ transform vectors $\bm{Z}^{m}_{o}$ and $\bm{D}^{m}$ back to decoded vectors $\Tilde{\bm{U}}^{m}_{o} , \Tilde{\bm{D}}^{m} \in \mathbb{R}^{\mathcal{C} \times d} $ respectively.

To realize cross-modal feature fusion, we employ a two-level reconstruction process with three networks $f^{va2t}(\cdot)$, $f^{ta2v}(\cdot)$ and $f^{tv2a}(\cdot)$ (detailed analysis in Appendix A). 
Taking the modality $t$ as an example, we first concatenate the intrinsic features $\Tilde{\bm{D}}^{t}$ and semantic features  $\Tilde{\bm{U}}^{\{v,a\}}_{o}$ in a certain modality order, where the former reflects the feature distribution of the current modality ($t$) and the latter provides semantic information of other modalities ($v,a$).
The concatenated vectors are input into $f^{va2t}(\cdot)$ to obtain the first-level reconstruction representations (FRR) $\bm{U}_{\alpha}^{t} \in \mathbb{R}^{\mathcal{C}\times d}$.
Then, $\bm{U}_{\alpha}^{m}$ of all modalities are concatenated and feed into $f^{va2t}(\cdot)$ to generate the second-level reconstruction representations (SRR) $\bm{U}_{\beta}^{t} \in \mathbb{R}^{\mathcal{C}\times d}$. The reconstruction-based fusion process of all modalities is expressed as,

\begin{figure*}[t]
\centering
\includegraphics[width=0.99 \linewidth]{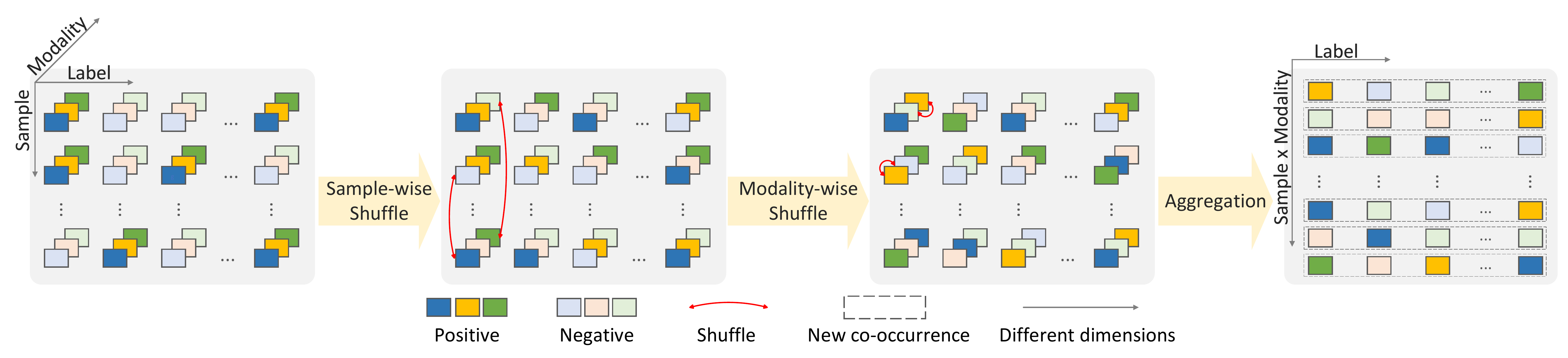} 
\caption{The semantic diagram of shuffle. The shuffle is conducted in both batch and modality dimensions. Different colors represent different modalities. After shuffling, new co-occurrence relations can appear in the training set (the gray dotted boxes).
}
\label{fig:method_shuffle}
\end{figure*}

\begin{equation}
\begin{aligned}
\bm{U}_{\alpha}^{t}\!&=\!f^{va2t}([\Tilde{\bm{D}}^{t};\Tilde{\bm{U}}^{v}_{o};\Tilde{\bm{U}}^{a}_{o}]), \! \bm{U}_{\beta}^{t}\!= \!f^{va2t}([\bm{U}_{\alpha}^{t};\bm{U}_{\alpha}^{v};\bm{U}_{\alpha}^{a}]), \\
\bm{U}_{\alpha}^{v}\!&=\!f^{ta2v}([\Tilde{\bm{U}}^{t}_{o};\Tilde{\bm{D}}^{v};\Tilde{\bm{U}}^{a}_{o}]), \!
\bm{U}_{\beta}^{v}\!= \!f^{ta2v}([\bm{U}_{\alpha}^{t};\bm{U}_{\alpha}^{v};\bm{U}_{\alpha}^{a}]), \\
\bm{U}_{\alpha}^{a}\!&=\!f^{tv2a}([\Tilde{\bm{U}}^{t}_{o};\Tilde{\bm{U}}^{v}_{o};\Tilde{\bm{D}}^{a}]), \!
\bm{U}_{\beta}^{a}\!=\! f^{tv2a}([\bm{U}_{\alpha}^{t};\bm{U}_{\alpha}^{v};\bm{U}_{\alpha}^{a}]).
\end{aligned}
\label{equ:two_level_reconstruction}
\end{equation}


To ensure that the reconstructed feature vectors can restore the original information,
we use the mean square error to formulate the reconstruction loss as:
\begin{equation}
\mathcal{L}_{rec} = \sum_{m}^{M}
\big(\|\bm{U}^{m}_{o}-\tilde{\bm{U}}^{m}_{o}\|_F + \|\bm{U}^{m}_{o}-\bm{U}_{\alpha}^m\|_F \big),
\label{equ:rec}
\end{equation}
where $\|\cdot\|_F$ returns the Frobenius norm of the matrix.

Due to the modality heterogeneity, different modalities express each emotion with different contributions.
Therefore, we introduce a Max Pooling-like network to impel each label to focus on its most relevant modality.
Specifically, we utilize three modality-specific classifiers $h_{\{t,v,a\}}(\cdot)$ on $\bm{U}^m_{o},\bm{U}^m_{\alpha},\bm{U}^m_{\beta}$ to calculate label prediction for $\{t,v,a\}$ modalities, respectively.
Then, we connect a Max-Pooling layer on these predictions to filter the most relevant modality of each label.
Taking $\bm{U}^{m}_{o}$ as an example, the final output via the above network is calculated as,
\begin{equation}
\bm{s}^o = \text{MaxPool}\big(h_{t}(\bm{U}^{t}_{o}), h_{v}(\bm{U}^{v}_{o}), h_{a}(\bm{U}^{a}_{o}) \big) \in \mathbb{R}^{\mathcal{C}}.
\label{equ:mp}
\end{equation}
In the same way, we can also obtain $\bm{s}^{\alpha}$ and $\bm{s}^{\beta}$.
Finally, we calculate the binary cross entropy (BCE) losses as,
\begin{equation}
\mathcal{L}_{cls}^{lsr} = 
\gamma_{o} l(\bm{s}^{o},\bm{y}) + \gamma_{\alpha} l(\bm{s}^{\alpha},\bm{y}) + \gamma_{\beta} l(\bm{s}^{\beta},\bm{y}),
\label{equ:lsr}
\end{equation}
where $l$ is the BCE loss and $\gamma_{o,\alpha,\beta}$ are trade-off parameters.

\subsubsection{\textbf{Contrastive Representation Learning.}}
To enable intrinsic vectors $\bm{D}^m$ to reflect the feature distribution of each label in different modalities,
we utilize contrastive learning to learn a distinguishable latent embedding space $\mathcal{S}^{z}$.
For samples in a batch of size $B$, after obtaining $\bm{U}^{m}_{o}$, $\bm{U}^{m}_{\alpha}$, $\bm{U}^{m}_{\beta}$, we feed them into the corresponding encoder $En^{m}(\cdot)$ to generate $L_2$-normalized latent embeddings $\bm{Z}^{m}_{o}$, $\bm{Z}^{m}_{\alpha}$, $\bm{Z}^{m}_{\beta} \in \mathbb{R}^{\mathcal{C} \times d_Z}$, respectively.
We follow the SCL \cite{khosla2020supervised} and additionally maintain a queue storing the most current latent embeddings, and we update the queue chronologically. 
Thus, we have the contrastive embedding pool as $\bm{E} = \{\bm{Z}^{\{t,v,a\}}_{\{o,\alpha,\beta\}}\}^{B}_{i=1} \cup \text{queue}$.
Given an anchor embedding $\bm{e} \in \bm{E}$, the contrastive loss is defined by contrasting its positive set with the remainder of the pool $\bm{E}$ as,
\begin{equation}
\mathcal{L}_{scl}(\bm{e},\bm{E})= - \frac{1}{|{\bm{P}}(\bm{e})|} \sum_{ \bm{e}^{+} \in \bm{P}(\bm{e})} \log \frac{\exp \left(  \bm{e}^{\top} \bm{e}^{+} / \tau\right)}{\sum\limits_{\bm{e}^{\prime} \in \bm{E}(\bm{e})} \exp \left(\bm{e}^{\top} \bm{e}^{\prime} / \tau\right)},
\end{equation}
where $\bm{P}(\bm{e})$ is the positive set and $\bm{E}(\bm{e})=\bm{E} \backslash \{\bm{e}\}$ . $\tau \in \mathbb{R}^{+}$ is the temperature.
The contrastive loss of the batch is:
\begin{equation}
\mathcal{L}_{scl}=\sum\nolimits_{\bm{e}\in\bm{E}}\mathcal{L}_{scl}(\bm{e},\bm{E}).
\label{equ:scl}
\end{equation}

To construct the positive set, 
considering the purpose of learning the modality-specific feature distribution of each label, we redefine the label for each $\bm{e}$.
According to the modality $m$, label category $j$ and label polarity $k$, the new label is defined as $\tilde{y}\!=\!l^{m}_{j,k}, m \!\!\in\! \{t,v,a\}, j\! \in\! [\mathcal{C}], k\! \in\! \{pos,neg\}$.
Thus, the positive examples are selected as $\bm{P}(\bm{e})\!=\!\{\bm{e}^{\prime}|\bm{e}^{\prime}\!\in\!\bm{E}(\bm{e}),\tilde{y}^{\prime}\!=\!\tilde{y}\}$, where $\tilde{y}^{\prime}$ is the label for $\bm{e}^{\prime}$.
In other words, the positive set is those embeddings from the same modality with the same label category and polarity.
Importantly, we keep a prototype embedding $\bm{\mu}^{m}_{j,k} \in \mathbb{R}^{d_z}$ corresponding to each class $l^{m}_{j,k}$, which can be deemed as a set of representative embedding vectors.
To reduce the computational toll and training latency, we update the class-conditional prototype vector in a moving-average style as,
\begin{equation}
\bm{\mu}^{m}_{j,k}=\text{Normalize}(\phi \bm{\mu}^{m}_{j,k} + (1-\phi)\bm{e}), \quad \text{if} \; \tilde{y}=l^{m}_{j,k},
\end{equation}
where the momentum prototype $\bm{\mu}^{m}_{j,k}$ is defined by the moving average of the normalized embedding whose defined class conforms to $l^{m}_{j,k}$. $\phi$ is a hyperparameter.
During training, we leverage prototypes to obtain the intrinsic vectors $\bm{D}^m = [\bm{d}_{1}^m,\dots, \bm{d}_{\mathcal{C}}^m]$ via the soft-max pattern as, 
\begin{equation}
\bm{d}_{j}^m=\sum_{k}^{\{pos,neg\}} \alpha^{m}_{j,k} \bm{u}_{j,k}^m, \; \alpha^{m}_{j,k}=\frac{\exp({{\bm{e}}^{\top} \bm{u}_{j,k}^m})}{\sum\limits_{k^{\prime}}^{\{pos,neg\}} \exp({{\bm{e}}^{\top} \bm{u}^m_{j,k^{\prime}}})},
\end{equation}
while during prediction, the hard-max pattern is used as,
\begin{equation}
\bm{d}_{j}^m\!= \!I_{\left[\alpha^{m}_{j,pos} \!>\!\alpha^{m}_{j,neg}\right]} \bm{u}_{j,pos}^m + I_{\left[\alpha^{m}_{j,pos} \!\leq\! \alpha^{m}_{j,neg}\right]} \bm{u}_{j,neg}^m, 
\end{equation}
where $I_{[\cdot]}$ is the indicator function.

\subsection{Shuffle-based Feature Aggregation}

Although exploiting the most relevant modality is sufficient to find discriminative features, multi-modal fusion can use complementary information to obtain more robust representations.
Therefore, we design a shuffle-based aggregation to exploit cross-modal information, which includes sample- and modality-wise shuffle processes.
The motivations of the sample- and modality-wise shuffle are to enrich the co-occurrence relations of labels and realize random cross-modal aggregation, respectively.
As shown in Figure \ref{fig:method_shuffle}, after obtaining the SRR of a batch of samples, two shuffle processes are performed sequentially and independently.
Specifically, we stack the vectors $\bm{U}_{\beta}^{m}$ of the batch as $\bm{V} = \text{Stack}\big(\big\{[\bm{U}_{\beta}^{t};\bm{U}_{\beta}^{v};\bm{U}_{\beta}^{a}]_{i}\big\}^{B}_{i=1}\big) \in \mathbb{R}^{B \times M \times \mathcal{C} \times d}$, where $M$ is the number of modalities.
Firstly, on each modality $m$, we perform the sample-wise shuffle (sws) as,
\begin{equation}
\bm{V}_{[:,m]}:=\left[\bm{v}_{1}^{m}, \dots, \bm{v}_{B}^{m}\right]
\stackrel{\text { sws }} {\rightarrow}
\tilde{\bm{V}}_{[:,m]}:=\left[\bm{v}_{r_1}^{m}, \dots, \bm{v}_{r_B}^{m}\right],
\end{equation}
where $\{r_{i}\}_1^B$ are new indices of samples.
Then, for each sample, the modality-wise shuffle (mws) is performed as,
\begin{equation}
\tilde{\bm{V}}_{[i,:]}:=\left[\tilde{\bm{v}}_{i}^{1}, \dots, \tilde{\bm{v}}_{i}^{M}\right]
\stackrel{\text { mws }} {\rightarrow}
\hat{\bm{V}}_{[i,:]}:=\left[\tilde{\bm{v}}_{i}^{r_{1}}, \dots, \tilde{\bm{v}}_{i}^{r_{M}}\right],
\end{equation}
where $\{r_{i}\}_1^M$ are new indices of modalities.
Then, $\bm{V}$ and $\hat{\bm{V}}$ are concatenated on the label dimension, respectively, as,
\begin{equation}
\begin{aligned}
\bm{Q} &= \Big\{ \big\{\bm{q}_i^m=[{\bm{v}}^{m}_{i,1}; \dots; {\bm{v}}^{m}_{i,\mathcal{C}}] \big\}^{\{t,v,a\}}_{m} \Big\}^{B}_{i} \in \mathbb{R} ^{B \times M \times \mathcal{C}\cdot d}, \\
    \hat{\bm{Q}} &= \Big\{ \big\{\hat{\bm{q}}_i^m=[\hat{\bm{v}}^{m}_{i,1}; \dots; \hat{\bm{v}}^{m}_{i,\mathcal{C}}] \big\}^{\{t,v,a\}}_{m} \Big\}^{B}_{i} \in \mathbb{R} ^{B \times M \times \mathcal{C}\cdot d}. \\
\end{aligned}
\end{equation}
It is worth noting that unlike $\bm{q}_i^m$, which is concatenated by features from a single modality and a single sample, the features constituting $\hat{\bm{q}}_i^m$ are randomly sampled from 1 to $M$ modalities and 1 to $\mathcal{C}$ samples.
Finally, the $\bm{Q}$ and $\hat{\bm{Q}}$ are used to fine-tune a classifier $h_{c}(\cdot)$ with the BCE loss as,
\begin{equation}
\mathcal{L}_{cls}^{agg} = \frac{1}{ M} \sum_{m}^{M}
\big(
l(h_c(\bm{q}^m),\bm{y}_{\bm{q}^m}) + \gamma_{sf}l(h_c(\hat{\bm{q}}^m),\bm{y}_{\hat{\bm{q}}^m})
\big),
\label{equ:agg}
\end{equation}
where $\gamma_{sf}$ is the trade-off parameter.
Combing the Equation \ref{equ:rec}, \ref{equ:lsr}, \ref{equ:scl} and \ref{equ:agg}, the final objective function is formulated as,
\begin{equation}
\mathcal{L} = \mathcal{L}_{cls}^{agg} + \mathcal{L}_{cls}^{lsr} +  + \gamma_{s}\mathcal{L}_{scl} + \gamma_{r}\mathcal{L}_{rec},
\end{equation}
where $\gamma_{s},\gamma_{r}$ are trade-off parameters.
During prediction, to utilize both the most relevant modality and multi-modal fusion, the prediction of the test sample $i^{\prime}$ is obtained as,
\begin{equation}
\hat{\bm{y}}_{i^{\prime}} = \frac{1}{2} (
\frac{1}{M} \sum\nolimits_{m}^{M} h_c(\bm{q}^{m}_{i^{\prime}}) + \bm{s}^{\beta }_{i^{\prime}}
) \in \mathbb{R} ^ {\mathcal{C}}.
\label{equ:prediction}
\end{equation}

\begin{table*}[!t]
\small
\centering
\begin{math}
\begin{array}{cc|cccc|cccc}
\hline 
\multirow{2}{*}{\text { Approaches }} & \multirow{2}{*}{\text { Methods }} & \multicolumn{4}{|c|}{\text { Aligned }} & \multicolumn{4}{c}{\text { Unaligned }} \\
\cline { 3 - 10 } & 
& \text { Acc } & \mathrm{P} & \mathrm{R} & \text { Micro-F1 } & \text { Acc } & \mathrm{P} & \mathrm{R} & \text { Micro-F1 } \\
\hline 
\multirow{3}{*}{\text { Classical }} & \text { BR } & 0.222 & 0.309 & 0.515 & 0.386 & 0.233 & 0.321 & 0.545 & 0.404 \\
& \text { LP } & 0.159 & 0.231 & 0.377 & 0.286 & 0.185 & 0.252 & 0.427 & 0.317 \\
& \text { CC } & 0.225 & 0.306 & 0.523 & 0.386 & 0.235 & 0.320 & 0.550 & 0.404 \\
\hline 
\multirow{3}{*}{\text { Deep-based }} & \text { SGM } & 0.455 & 0.595 & 0.467 & 0.523 & 0.449 & 0.584 & 0.476 & 0.524 \\
& \text { LSAN } & 0.393 & 0.550 & 0.459 & 0.501 & 0.403 & 0.582 & 0.460 & 0.514 \\
& \text { ML-GCN } & 0.411 & 0.546 & 0.476 & 0.509 & 0.437 & 0.573 & 0.482 & 0.524 \\
\hline 
\multirow{7}{*}{\text { Multi-modal }} & \text { MulT } & 0.445 & 0.619 & 0.465 & 0.531 & 0.423 & 0.636 & 0.445 & 0.523 \\
& \text { MISA } & 0.430 & 0.453 & \bm{0 . 5 8 2} & 0.509 & 0.398 & 0.371 & \bm{0 . 5 7 1} & 0.450 \\
& \text { MMS2S } & 0.475 & 0.629 & 0.504 & 0.560 & 0.447 & 0.619 & 0.462 & 0.529 \\
& \text { HHMPN } & 0.459 & 0.602 & 0.496 & 0.556 & 0.434 & 0.591 & 0.476 & 0.528 \\
& \text { TAILOR } & 0.488 & 0.641 & 0.512 & 0.569 & 0.460 & 0.639 & 0.452 & 0.529 \\
& \text{ AMP } & 0.484 & 0.643 & 0.511 & 0.569 & 0.462 & 0.642 & 0.459 & 0.535 \\
\cline { 2 - 10 } &
\text{ CARAT } & \bm{0.494} & \bm{0.661} & 0.518 & \bm{0.581} & \bm{0.466} & \bm{0.652} & 0.466 & \bm{0.544} \\

\hline
\end{array}
\end{math}
\caption{ 
Performance comparison between CARAT and baselines on CMU-MOSEI dataset with aligned and unaligned settings. 
}
\label{tab:result_main}
\end{table*}

\section{Experiments}

\subsection{Experimental Settings}
\subsubsection{\textbf{Dataset and Evaluation Metrics.}}
We evaluate CARAT on two benchmark MMER datasets (CMU-MOSEI \cite{zadeh2018multimodal} and M${}^{3}$ED \cite{zhao2022m3ed}), which maintained settings in the public SDK\footnote{\url{https://github.com/A2Zadeh/CMU-MultimodalSDK} \label{ft:CMU}}\footnote{\url{https://github.com/AIM3-RUC/RUCM3ED} \label{ft:m3ed}} .
Four evaluation metrics are employed: Accuracy (Acc), Micro-F1, Precision (P), and Recall (R). 
More detailed descriptions and preprocessing of datasets are shown in Appendix B.

\subsubsection{\textbf{Baselines.}}

We compare CARAT with various approaches of two groups. 
The first group is Multi-Label Classification (MLC) methods.
Specifically,
in these approaches, the multi-modal inputs are
early fused (simply concatenated) as a new input.
For classic methods:
(1) \textbf{BR} 
\cite{boutell2004learning} transforms MLC into multiple binary classifications while ignoring label correlations.
(2) \textbf{LP} 
\cite{tsoumakas2006multi} breaks the initial label set into several random subsets and trains a corresponding classifier.
(3) \textbf{CC} 
\cite{read2011classifier} transforms MLC into a chain of binary classification problems by considering high-order label correlations.
For deep-based methods:
(4) \textbf{SGM} 
\cite{yang2018sgm} views MLC as a sequence generation problem via label correlation.
(5) \textbf{LSAN} 
\cite{xiao2019label} explores the semantic connection between labels and documents to construct label-specific document representation.
(6) \textbf{ML-GCN} 
\cite{chen2019multi} employs GCN to map label representations and captures label correlations for image recognition.

The second group is multi-modal multi-label methods.
(7) \textbf{MulT} 
\cite{tsai2019multimodal} uses cross-modal interactions to fuse information from one modality to another. 
(8) \textbf{MISA} 
\cite{hazarika2020misa} learns modality-invariant and modality-specific representations for the fusion.
(9) \textbf{MMS2S} \cite{zhang2020multi}
handles the modality and label dependence in a sequence-to-set approach.
(10) \textbf{HHMPN} 
\cite{zhang2021multi} models feature-to-label, label-to-label and modality-to-label dependencies via graph message passing.
(11) \textbf{TAILOR}
\cite{zhang2022tailor} adversarially depicts commonality and diversity among modalities to obtain discriminative representations.
(12) \textbf{AMP} \cite{ge2023learning} learns robust representations with adversarial temporal masking and Adversarial Parameter Perturbation.

\subsubsection{\textbf{Implementation Details.}}
We set the size of hidden states as $d\!=\!256, d_z\!=\!64$.
The size of the embedding queue is set to 8192.
All encoders $En^m(\cdot)$ and decoders $De^m(\cdot)$ are implemented by 2-layer MLPs.
We set hyper-parameters $\gamma_{o}\!=\!0.01$, $\gamma_{\alpha}\!=\!0.1$, $\gamma_{\beta}\!=\!1$, $\gamma_{s}\!=\!1$, $\gamma_{sf}\!=\!0.1$ and $\gamma_{r}\!=\!1$ and the analysis of different weight settings is presented in Appendix A. 
We set $l_t\!=\!6,l_v\!=\!l_a\!=\!4$ for the layer number of Transformer Encoders.
We employ the Adam \cite{kingma2014adam} optimizer with the initial learning rate of $5e^{-5}$ and a liner decay learning rate schedule with a warm-up strategy.
The batch size $B$ is set to 64.
During training, we train methods for 20 epochs to select the model with the best F1 score on the validation set as our final model.
All experiments are conducted with one NVIDIA A100 GPU.

\subsection{Experimental Results}

\subsubsection{\textbf{Performance Comparison.}}
We show performance comparisons on CMU-MOSEI and M${}^{3}$ED (only partial multi-modal baselines) in Table \ref{tab:result_main},\ref{tab:m3ed_res},
and observations are as follows:
1) CARAT significantly outperforms all rivals by a significant margin.
Although \textbf{MISA} has a prominent recall, its precision drops to a poor value and its performance is far inferior to CARAT on more important metrics (Micro-F1 and accuracy).
Furthermore, 
CARAT still maintains a decent performance boost in the unaligned setting,
which proves that CARAT can break the barrier of the modality gap better than others.
2) Among uni-modal approaches, the superior performance of deep-based methods, i.e. \textbf{SGM}, \textbf{LSAN} and \textbf{ML-GCN}, over classic methods, i.e. \textbf{BR}, \textbf{CC} and \textbf{LP}, indicates that deep representation can better capture semantic features and label correlations help to capture more meaningful information.
3) Compared with uni-modal approaches, multi-modal methods typically exhibit better performance, which shows the necessity of modeling multi-modal interactions.
4) Among all baselines, \textbf{TAILOR} achieves competitive performance, which validates the effectiveness of leveraging commonality and diversity among modalities to obtain the discriminative label representations.

\begin{table}[!t]
\small
\centering
\begin{math}
\begin{array}{l|cccc}
\hline \text { Methods } & \text { Acc } & \text { P } & \text { R } & \text { Micro-F1 } \\
\hline  
\text { MMS2S}  & 0.645 & 0.813 & 0.737 & 0.773 \\
\text { HHMPN } & 0.648 & 0.816 & 0.743 & 0.778 \\
\text { TAILOR}  & 0.647 & 0.814 & 0.739 & 0.775 \\
\text { AMP}  & 0.654 & 0.819 & 0.748 & 0.782 \\
\hline
\text { CARAT } & \mathbf{0.664} & \mathbf{0.824} & \mathbf{0.755} & \mathbf{0.788} \\
\hline
\end{array}
\end{math}
\caption{
Performance comparison on the M${}^{3}$ED dataset.
}
\label{tab:m3ed_res}
\end{table}

\begin{table}[!t]
\small
\centering
\begin{math}
\begin{array}{l|cccc}
\hline \text { Approaches } & \text { Acc } & \text { P } & \text { R } & \text {Micro-F1} \\
\hline \text { (1) MRM + AGG } & 0.475 & 0.647 & 0.507 & 0.569 \\
\text { (2) only MRM } & 0.474 & 0.641 & 0.502 & 0.563 \\
\text { (3) only AGG } & 0.472 & 0.639 & 0.506 & 0.565 \\
\hline 
\text { (4) w/o } \mathcal{L}_{scl} & 0.481 & 0.640 & 0.515 & 0.571 \\
\text { (5) w/o } En,De & 0.475 & 0.638 & 0.514 & 0.569 \\
\hline 
\text { (6) w/o } \mathcal{L}_{rec} & 0.482 & 0.644 & 0.516 & 0.573 \\
\text { (7) w/o } \alpha \text{-recon} & 0.483 & 0.636 & \mathbf{0.520} & 0.572 \\
\text { (8) w/o } \beta \text{-recon} & 0.482 & 0.631 & 0.513 & 0.566 \\
\text { (9) w/o } \alpha\&\beta \text{-recon} & 0.475 & 0.619 & 0.503 & 0.555 \\
\hline 

\text { (10) w/o sw-shf}  & 0.491 & 0.659 & 0.511 & 0.575 \\
\text { (11) w/o mw-shf } & 0.490 & 0.656 & 0.514 & 0.576 \\
\text { (12) w/o shf}  & 0.489 & 0.658 & 0.509 & 0.574 \\
\hline
\text { (13) CARAT } & \mathbf{0.494} & \mathbf{0.661} & 0.518 & \mathbf{0.581} \\
\hline
\end{array}
\end{math}
\caption{ 
Ablation tests on the aligned CMU-MOSEI.
"MRM" and "AGG" respectively denote using features of the Most Relevant Modality and AGGregated features.
"w/o" means removing.
"w/o $En,De$" denotes removing the encoding and decoding.
"w/o $\{ \alpha, \beta, \alpha\&\beta \}$-recon" denotes removing the first-level reconstruction, second-level reconstruction, or both.
"w/o $\{$ sw-, mw-, $\phi\}$shf" denotes removing the sample-, modality-wise shuffle process or both.
Detailed implementations are in Appendix A.
}
\label{tab:result_ablation}
\end{table}

\begin{figure*}[t]
\centering
\begin{subfigure}{0.14\linewidth}
    \centering
    \includegraphics[width=0.99\linewidth]{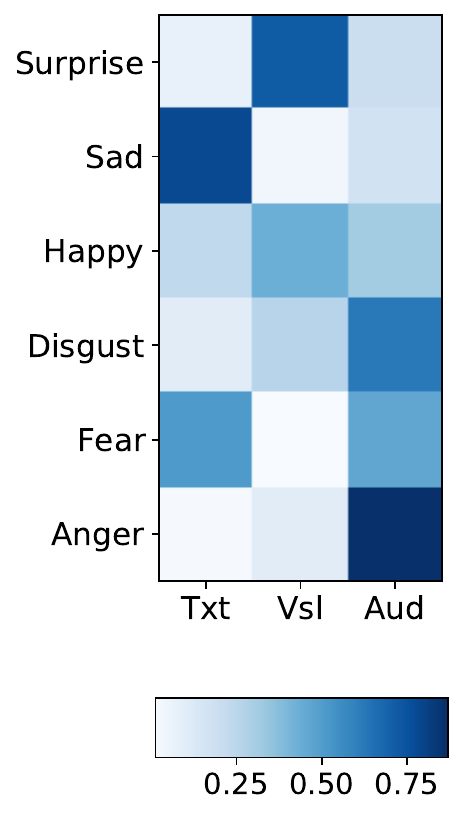}
\caption{ }
\end{subfigure}
\begin{subfigure}{0.85\linewidth}
    \centering
    \includegraphics[width=1.0\linewidth]{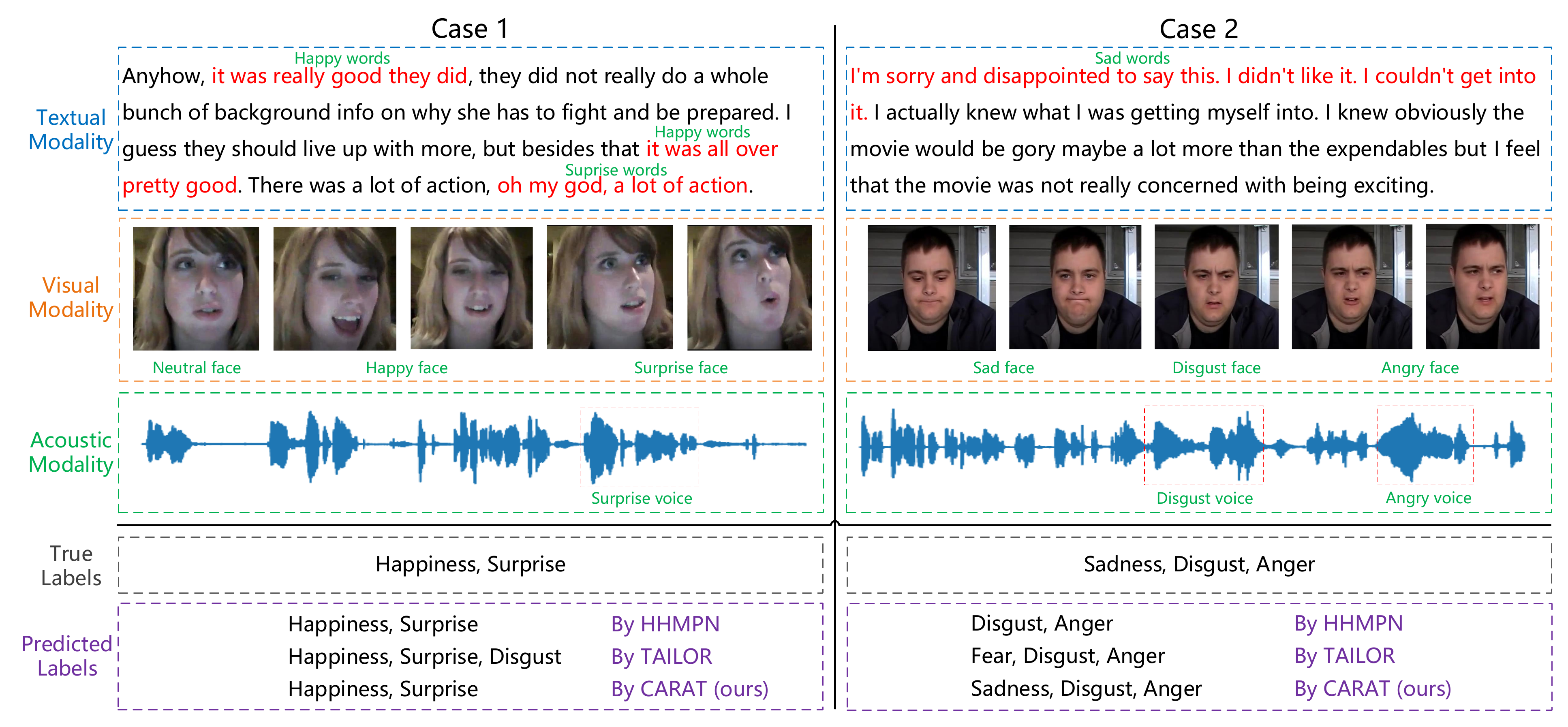}
\caption{ }
\end{subfigure}
\caption{
(a) The visualization of modality-to-label dependencies, indicating the correlation of labels in each row to modality in each column, where darker colors indicate stronger correlations. (b) Two cases of emotion recognition by multiple methods.
}
\label{fig:exp_m2l_freq}
\end{figure*}

\subsubsection{\textbf{Ablation Study.}}
To demonstrate the importance of each component, we compare CARAT with various ablated variants.
As shown in Table \ref{tab:result_ablation}, we can see:

1) \emph{Effect of exploiting both specificity and complementarity:}
By using both features of the most relevant modality (MRM) and aggregated features (AGG), (1) is better than (2) and (3), which indicates the significance of binding modality specificity and complementarity.

2) \emph{Effect of the contrastive representation learning:}
Without conducting the loss $\mathcal{L}_{scl}$, (4) is worse than CARAT, which illustrates the significance of leveraging contrastive learning to learn distinguishable representations.
Further, by removing the process of encoding and decoding,  
(5) is worse than (4), which validates the rationality of exploring the intrinsic embeddings in the latent space.

3) \emph{Effect of the two-level feature reconstruction:}
First, (6) is worse than CARAT, which reveals the effectiveness of using loss $\mathcal{L}_{rec}$ to constrain feature reconstruction.
Removing the first- and second-level reconstruction processes, (7) and (8) have different degrees of performance degradation compared to CARAT.
When the entire reconstruction process is removed, the performance of (9) is further reduced than (7) and (8), which confirms the effectiveness of multi-level feature reconstruction to achieve multi-modal fusion.

4) \emph{Effect of different shuffling operations:}
Excluding any round of the shuffling process, (10) and (11) are worse than CARAT, and (12) is even worse when both shuffling processes are removed, which confirms the effectiveness of performing shuffling in both sample and modality dimensions.


\begin{figure}[ht]
\centering
	\begin{subfigure}{0.495\linewidth}
		\centering
		\includegraphics[width=1.0\linewidth]{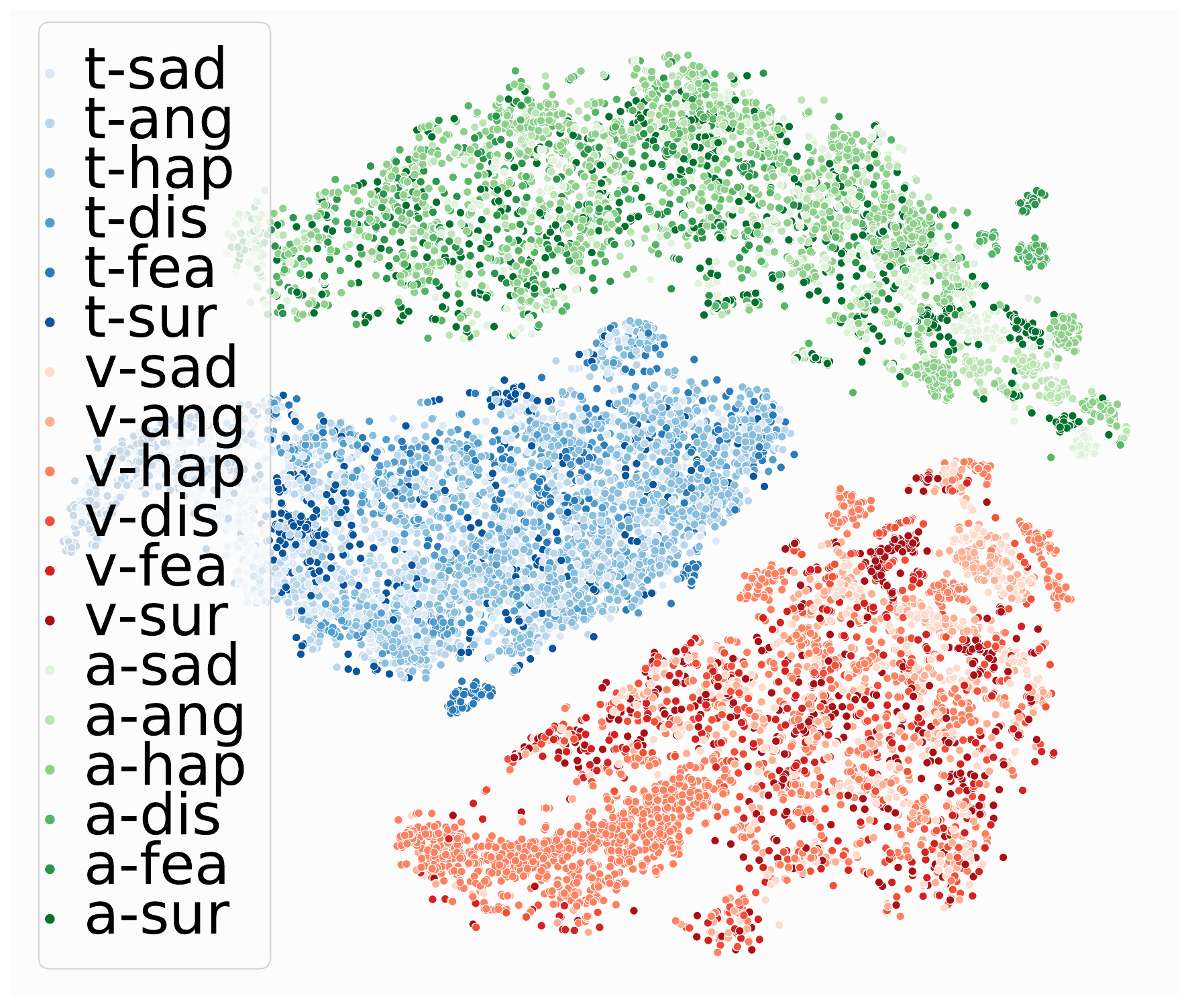}
	\end{subfigure}
	\begin{subfigure}{0.495\linewidth}
		\centering
		\includegraphics[width=1.0\linewidth]{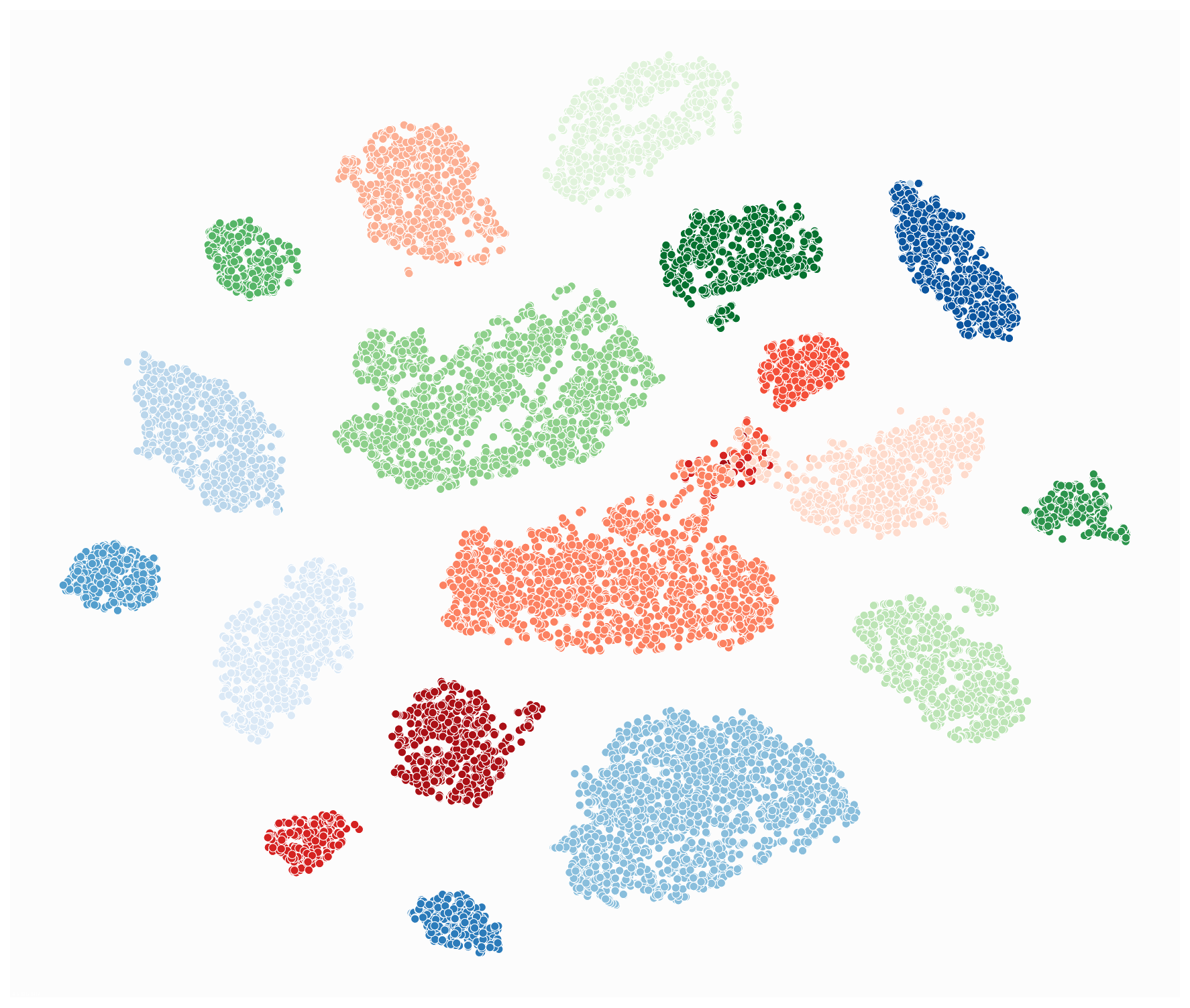}
	\end{subfigure}    
\caption{t-SNE visualization of embeddings without/with (left/right) CL. 
Different colors represent different modalities and different shades represent different emotions.
}

\label{fig:exp_latent}
\end{figure}


\subsection{Analysis}

\subsubsection{\textbf{Visualization of Distinguishable Representations.}}
To investigate the efficacy of contrastive learning (CL) on distinguishable representations, we visualize embeddings $\bm{Z}_o^m$ in space $\mathcal{S}^{z}$ using t-SNE \cite{van2008visualizing} without or with contrastive learning on the aligned CMU-MOSEI dataset.
As shown in Figure \ref{fig:exp_latent}, without CL (left subfigure), although embeddings belonging to different modalities can be well distinguished, the embeddings of different classes in each modality are lumped together.
In contrast, in the right subfigure, embeddings belonging to different modalities and labels are explicitly separated from each other, and embeddings of the same modality still maintain a tighter distribution.
Thus, with CL, CARAT produces well-separated clusters and more distinguishable representations.

\subsubsection{\textbf{Visualization of Modality-to-label Correlations.}}
We visualize the correlation of labels with their most relevant modalities.
As shown in Figure \ref{fig:exp_m2l_freq} (a),
each label focuses on different modalities unequally and usually has its own prone modality.
The modality-to-label correlations differ from label to label, 
e.g., emotion \emph{surprise} and \emph{sad} are highly correlated with the visual and textual modality, respectively.
More visualization examples are shown in Appendix E.

\subsubsection{\textbf{Case Study.}}
To further demonstrate the effectiveness of CARAT, Figure \ref{fig:exp_m2l_freq} (b) presents two cases.
1) Emotions expressed by different modalities are not consistent, which reflects the modality specificity.
E.g., in Case 2, emotion \emph{angry} can be mined intuitively from the visual and audio, but not the text.
2) HHMPN wrongly omitted relevant labels due to neglecting modality specificity, which results in the inability to capture richer semantic information.
In contrast, TAILOR gives wrong related labels.
Since TAILOR uses self-attention that can only explore label correlations within each sample, global information cannot be exploited.
Overall, our CARAT achieves the best performance.

\section{Conclusion}
In this paper, we propose ContrAstive feature Reconstruction and AggregaTion (CARAT) for MMER, which integrates effective representation learning and multiple dependency modeling into a unified framework.
We propose a reconstruction-based fusion mechanism by contrastively learning modal-separated and label-specific features to model fine-grained modality-to-label dependencies.
To further exploit the modality complementarity, we introduce a shuffle-based aggregation strategy to enrich co-occurrence collaboration among labels.
Experiments on benchmark datasets CMU-MOSEI and M${}^{3}$ED demonstrate the effectiveness of CARAT over state-of-the-art methods.

\section{Acknowledgments}
This work is supported by the National Key R\&D Program of China (No.2022YFB3304100) and the Pioneer R\&D Program of Zhejiang "Data Preparation, Optimization, and Augmentation for Domain-Specific Large Models".

\bibliography{aaai24}

\appendix
\clearpage

\begin{table*}[t]
\small
\centering
\begin{math}
\begin{array}{c|cccccc|cccc}
\hline 
\text { ID } & \; {\gamma}_o \; & \; {\gamma}_{\alpha} \; & \; {\gamma}_{\beta} \; & \; {\gamma}_{sf} \; & \; {\gamma}_{s} \; & \; {\gamma}_{r} \; & \text { Micro-F1 } & \text { P } & \text { R } & \text { Acc } \\
\hline
0 & 0.01 & 0.1 & 1 & 0.1 & 1 & 1 & \bm{0.581} & \bm{0.661} & \bm{0.518} & \bm{0.494} \\
\hline
1 & \underline{0} & 0.1 & 1 & 0.1 & 1 & 1 & 0.577 & 0.657 & 0.514 & 0.491 \\
2 & \underline{0.001} & 0.1 & 1 & 0.1 & 1 & 1 & 0.578 & 0.658 & 0.516 & 0.492 \\
3 & \underline{0.1} & 0.1 & 1 & 0.1 & 1 & 1 & 0.578 & 0.657 & 0.516 & 0.492 \\
\hline
4 & 0.01 & \underline{0} & 1 & 0.1 & 1 & 1 & 0.576 & 0.658 & 0.512 & 0.489 \\
5 & 0.01 & \underline{0.01} & 1 & 0.1 & 1 & 1 & 0.577 & 0.659 & 0.513 & 0.490 \\
6 & 0.01 & \underline{1} & 1 & 0.1 & 1 & 1 & 0.576 & 0.657 & 0.513 & 0.489 \\
\hline
7 & 0.01 & 0.1 & \underline{0} & 0.1 & 1 & 1 & 0.560 & 0.650 & 0.492 & 0.483 \\
8 & 0.01 & 0.1 & \underline{0.1} & 0.1 & 1 & 1 & 0.572 & 0.653 & 0.509 & 0.491 \\
9 & 0.01 & 0.1 & \underline{2} & 0.1 & 1 & 1 & 0.578 & 0.661 & 0.513 & 0.488 \\
\hline
10 & 0.01 & 0.1 & 1 & \underline{0} & 1 & 1 & 0.574 & 0.658 & 0.509 & 0.489 \\
11 & 0.01 & 0.1 & 1 & \underline{0.01} & 1 & 1 & 0.576 & 0.654 & 0.514 & 0.492 \\
12 & 0.01 & 0.1 & 1 & \underline{1} & 1 & 1 & 0.577 & 0.655 & 0.516 & 0.490 \\
\hline
13 & 0.01 & 0.1 & 1 & 0.1 & \underline{0} & 1 & 0.571 & 0.640 & 0.515 & 0.481 \\
14 & 0.01 & 0.1 & 1 & 0.1 & \underline{0.1} & 1 & 0.574 & 0.652 & 0.513 & 0.488 \\
15 & 0.01 & 0.1 & 1 & 0.1 & \underline{2} & 1 & 0.573 & 0.651 & 0.512 & 0.488 \\
\hline
16 & 0.01 & 0.1 & 1 & 0.1 & 1 & \underline{0} & 0.573 & 0.644 & 0.516 & 0.482 \\
17 & 0.01 & 0.1 & 1 & 0.1 & 1 & \underline{0.1} & 0.574 & 0.652 & 0.513 & 0.487 \\
18 & 0.01 & 0.1 & 1 & 0.1 & 1 & \underline{2} & 0.572 & 0.657 & 0.507 & 0.488 \\
\hline
\end{array}
\end{math}
\caption{
Performance comparison of different hyper-parameter settings on the aligned CMU-MOSEI.
ID 0 is the final hyper-parameter setting applied in our CARAT.
}
\label{tab:loss_hyper} 
\end{table*}

\section{A. Analysis of the Reconstruction-based Fusion Mechanism}
\label{app:exp_setting}

\subsection{A.1 Further Explanation of the Two-level Feature Reconstruction}

\subsubsection{Why reconstruction is set to two-level?}

First of all, the proposed feature reconstruction operation is a novel multi-modal fusion strategy different from the aggregation- and alignment-based methods, which intends to utilize the feature distribution information and the information of other modalities to restore the semantic features of the current modality. Experimental results show the effectiveness of the proposed method. The proposed reconstruction-based strategy ensures that the model can better capture the unique feature distribution information of each modality, while also implicitly integrating key semantic information of other modalities. The former is the focus of the first-level reconstruction process, while the latter is the focus of the second-level reconstruction process.

Compared with the one-level reconstruction process, we believe that the two-level reconstruction process can better learn the mutual reconstruction process across multiple modalities. In the first-level reconstruction process, we use the feature distribution information learned from the latent space and the semantic information of other modalities to restore the representation of the current modality to preserve the characteristic information of each modality. To obtain a better label-specific representation, we add the second-level reconstruction process to further strengthen the learning of the reconstruction network. This stacked reconstruction process is inspired by the design of stacked deep neural networks. Experimentally, the ablation study strongly confirmed the effectiveness of the devised two-level reconstruction process.

\subsubsection{What's the essential difference between the two reconstruction features $\bm{U}_{\alpha}^{t,v,a}$ and $\bm{U}_{\beta}^{t,v,a}$?}

First of all, we want to explain that the learning process of the three reconstruction networks is mainly affected by $\mathcal{L}_{rec}$ and $\mathcal{L}_{cls}^{lsr}$. 
$\mathcal{L}_{rec}$ is used to ensure that the reconstructed representation is close to the original label-specific representation, thus preserving modality specificity. 
$\mathcal{L}_{cls}^{lsr}$ is to introduce supervised information so that the reconstructed representation has the correct label polarity, which can be used for label correlation prediction. 
$\mathcal{L}_{rec}$ is only implemented in the first-level reconstruction process, to strengthen the retention of the information of the respective modality characteristics and weaken the role of cross-modal fusion. In the second-level reconstruction process, without the constraints of $\mathcal{L}_{rec}$, 
$\bm{U}_{\beta}^{t,v, a}$ can fuse information from other modalities to a relatively large extent. However, since the three reconstructed networks have been constrained to restore the representation of the corresponding modality in the first-level reconstruction process, the three networks can still maintain this property in the second-level reconstruction process, but the purpose of implicit multi-modal fusion is amplified. As a result, the reconstructed $\bm{U}_{\alpha}^{t,v,a}$ and $\bm{U}_{\beta}^{t,v,a}$ have different characteristics and usage purposes.

This explains why Eq. \ref{equ:prediction} uses $\bm{s}^{\beta}$ and not $\bm{s}^{\alpha}$.
During the training process, utilizing $\bm{s}^{\alpha}$ to calculate BCE loss is to better constrain $\bm{U}_{\alpha}^{t,v, a}$ to maintain the correct label polarity, which is beneficial to downstream learning. In the prediction process, only using $\bm{s}^{\beta}$ is based on two considerations: First, as answered in Question 2.2), we believe that through the two-level feature reconstruction operation, the deeper $\bm{U}_{\beta}^{t,v, a}$has better semantic representation than $\bm{U}_{\alpha}^{t,v, a}$, which can lead to better prediction performance. Second, because the learning of $\bm{U}_{\alpha}^{t,v,a}$ is also constrained by the loss $\mathcal{L}_{rec}$ during the training process, the prediction performance of $\bm{s}^{\alpha}$ calculated by $\bm{U}_{\alpha}^{t,v,a}$ will be affected to a certain extent.

\subsection{A.2 Analysis of the Parameters}
To better explain the reasons for setting different weights for each parameter in the training loss, we give more experimental results to show the impact of setting different values for them. Experiments are implemented on the CMU-MOSEI dataset of the data-aligned setting. As shown in Table \ref{tab:loss_hyper}, ID 0 is the final hyperparameter setting used in our paper, which is the baseline for comparison with other settings. Compared with ID 0, ID 1-18 only changes the weight of a certain hyperparameter.

(1) First of all, it can be seen that ID 0 maintains optimal performance, and other settings have a certain degree of performance degradation.

(2) For each hyperparameter, the greater the difference from the preset value in ID 0, the greater the degree of performance degradation, which can prove that the value set in ID 0 is relatively reasonable.

(3) Comparing the experimental results of ID 1-9, it can be found that the change of ${\gamma}_{o}$, ${\gamma}_{\alpha}$, ${\gamma}_{\beta}$ values gradually increases the impact on the results, which indicates that the three have different importance, and also confirms the reason why we use different weights for them.

(4) The experimental results on ID 10-18 also show that ${\gamma}_{sf}$, ${\gamma}_{s}$, ${\gamma}_{r}$ have different effects on performance. When the weight is set to 0, the large performance gap justifies the use of corresponding losses.

\section{B. Dataset}
\label{app:dataset}

Here we give an additional description of two datasets.

\subsection{B.1 CMU-MOSEI}
The dataset contains 22,856 utterance-level video clips segmented from 3,229 full-long videos with 1000 distinct speakers.
Each video clip is annotated with multiple labels from 6 discrete emotions $\{ fear$, $ happiness, sadness, anger, disgust,$ $ surprise \}$,
according to three modalities, i.e., the textual, visual, and acoustic modalities.
The average words of video clips are 19.1 and the average number of emotion labels per sample is 1.6.
Table \ref{tab:dataset_sta} summarizes the statistics of the samples with multiple labels.
The training, validation, and test data are all the same size as the video clips in the public SDK\footnote{\url{https://github.com/A2Zadeh/CMU-MultimodalSDK}}, with approximate size 16.3K, 1.9K, and 4.7K, respectively.
Follow \cite{zhang2022tailor,ju2020transformer}, we utilize 300-dimensional text features from manual transcripts by the GloVe word embeddings \cite{pennington2014glove}, 35-dimensional visual features from video frames by the library FACET \cite{baltruvsaitis2016openface} and 74-dimensional acoustic features from acoustic signals by the COVAREP software \cite{degottex2014covarep}. 
We implement the word-level alignment for the experiments on aligned data and retain the originally extracted features for the experiments on unaligned data.
Table \ref{tab:dataset_split} shows details of CMU-MOSEI in both data-aligned and unaligned settings.

\begin{table}[h]
\small
\centering
\begin{math}
\begin{array}{c|c|c|c}
\hline \text { Multi-label } & \text { Number } & \text { Emotion } & \text { Number } \\
\hline \text { None } & 3372 & \text { Happiness } & 12240 \\
\hline \text { One } & 11050 & \text { Surprise } & 1892 \\
\hline \text { Two } & 5526 & \text { Sadness } & 5918 \\
\hline \text { Three } & 2084 & \text { Anger } & 4933 \\
\hline \text { Four } & 553 & \text { Disgust } & 3680 \\
\hline \text { Five } & 84 & \text { Fear } & 2286 \\
\hline \text { Six } & 8 & - & - \\
\hline
\end{array}
\end{math}
\caption{The emotion statistics on the CMU-MOSEI dataset.}
\label{tab:dataset_sta} 
\end{table}


\begin{table}[h]
\small
\centering
\begin{math}
\begin{array}{c|cc|cc|cc}
\hline 
\multirow{3}{*}{aligned} & N_{train} & 16,326 & d_v & 35 & n_v & 60 \\
& N_{valid} & 1,871 & d_a & 74 & n_a & 60 \\
& N_{test} & 4,659 & d_t & 300 & n_t & 60 \\
\hline
\multirow{3}{*}{unaligned} & N_{train} & 16,326 & d_v & 35 & n_v & 500 \\
& N_{valid} & 1,871 & d_a & 74 & n_a & 500 \\
& N_{test} & 4,659 & d_t & 300 & n_t & 50 \\
\hline 
\end{array}
\end{math}
\caption{Statistics of CMU-MOSEI. $N_{\{train, valid, test\}}$ is the size of training, validation and test sets, $d_{\{v,a,t\}}$ is modality dimension and $n_{\{v,a,t\}}$ is sequence length.}
\label{tab:dataset_split}
\end{table}

\subsection{B.2 M${}^3$ED}

M${}^3$ED is a large-scale Multi-modal Multi-scene and Multi-label Emotional Dialogue
dataset. 
Considering that M${}^3$ED is a dialog corpus, each dialog instance is composed of multiple single-label sentences. But the entire dialogue can be regarded as a multi-label instance, which can be applied to the MMER task. In addition, M${}^3$ED is a data-aligned dataset.

M${}^3$ED contains 990 dialogues, 9,082 turns, 24,449 utterances derived from 56 different TV series (about 500 episodes), which ensures the scale and diversity of the dataset.
M${}^3$ED adopts the TV-independent data split manner in order to avoid
any TV-dependent bias, which means there is no overlap of TV series across training, validation, and testing sets.
The basic statistics are similar across these three data splits. There are rich emotional interaction phenomena in the M${}^3$ED, for example, 5,396 and 2,696 inter-turn emotion-shift and emotion-inertia scenarios respectively, and 2,879
and 10,891 intra-turn emotion-shift and emotioninertia scenarios.
Table \ref{tab:dataset_m3ed_dis} presents the single emotion distribution statistics. The distribution of each emotion category is similar across train/val/test sets.

\begin{table}[h]
\small
\centering
\begin{math}
\begin{array}{ccccc}
\hline \text { Emotion } & \text { Train } & \text { Val } & \text { Test } & \text { Total } \\
\hline 
\text { neutral } & 7,130 & 1,043 & 1,855 & 10,028 \\
\text { happy } & 1,626 & 303 & 358 & 2,287 \\
\text { surprise } & 696 & 120 & 235 & 1,051 \\
\text { sad } & 2,734 & 489 & 734 & 3,957 \\
\text { disgust } & 1,145 & 134 & 218 & 1,497 \\
\text { anger } & 3,816 & 682 & 736 & 5,234 \\
\text { fear } & 280 & 50 & 65 & 395 \\
\hline \text { Total } & 17,427 & 2,821 & 4,201 & 24,449 \\
\hline
\end{array}
\end{math}
\caption{Emotion Distribution of M${}^3$ED.}
\label{tab:dataset_m3ed_dis}
\end{table}

\section{C. Implementation of Ablation Variants}

\subsection{C.1 The key details of MRM+AGG}

"MRM" and "AGG" respectively denote using features of the most relevant modality and aggregated features. Their specific implementation methods are as follows: Firstly, both of them obtain label-specific representations $\bm{U}^{t,v, a}_{o}$ via uni-modal label-specific feature extraction, but remove the following two steps in CARAT, i.e. contrastive multi-modal feature reconstruction and shuffle-based feature aggregation. MRM connects a Max Pooling-like network to discover the representation in the most relevant modality of each label to measure the correlation. AGG directly concatenates label-specific representations of multiple modalities for each label and then uses the aggregated features to calculate the correlation. MRM+AGG uses both MRM and AGG to obtain a more comprehensive correlation prediction.

\subsection{C.2 The influence of removing $En$ and $De$ on the following shuffle and prediction process}

Removing $En$ and $De$ means removing the encoding and decoding of label-specific representations and the process of contrastive learning. This operation will only affect the first-level feature reconstruction process, where Eq. \ref{equ:two_level_reconstruction} is partially replaced with,
\begin{equation}
\begin{aligned}
    \bm{U}_\alpha^t=f^{v a 2 t}\left(\left[\bm{U}^t_o ; \bm{U}^v_o ; \bm{U}^a_o\right]\right), \\
    \bm{U}_\alpha^v=f^{t a 2 v}\left(\left[\bm{U}^t_o ; \bm{U}^v_o ; \bm{U}^a_o\right]\right), \\
    \bm{U}_\alpha^a=f^{t v 2 a}\left(\left[\bm{U}^t_o ; \bm{U}^v_o ; \bm{U}^a_o\right]\right) .
\end{aligned}
\end{equation}
The calculation flow from the second-level feature reconstruction to the end will not change.

\begin{table}[!t]
\small
\centering
\begin{math}
\begin{array}{l|cccc}
\hline 
\text { Approaches } & \text { Acc } & \text { P } & \text { R } & \text {Micro-F1} \\
\hline 
\text {alignment-based} & 0.469 & 0.622 & 0.498 & 0.553 \\
\text {aggregation-based} & 0.471 & 0.629 & 0.502 & 0.558 \\
\text {reconstruction-based} & \bm{0.473} & \bm{0.632} & \bm{0.504} & \bm{0.560} \\
\hline 
\end{array}
\end{math}
\caption{ 
Performance comparison of different multi-modal fusions on the aligned CMU-MOSEI dataset.
}
\label{tab:result_fusion}
\end{table}

\section{D. Comparison of Different Fusion}
\label{app:fusion}
To demonstrate the effectiveness of reconstruction-based fusion, we compare it with alignment- and aggregation-based fusions.
To highlight the performance difference caused by different fusion methods, we simplify CARAT by only retaining the extraction and fusion components.
As shown in Table \ref{tab:result_fusion}, the superior performance of (3) over (1) and (2) illustrates the effectiveness of reconstruction-based fusion that considers modality complementarity and specificity.

\begin{figure}[t]


\begin{subfigure}{0.99\linewidth}
    \centering
    \includegraphics[width=0.9\linewidth]{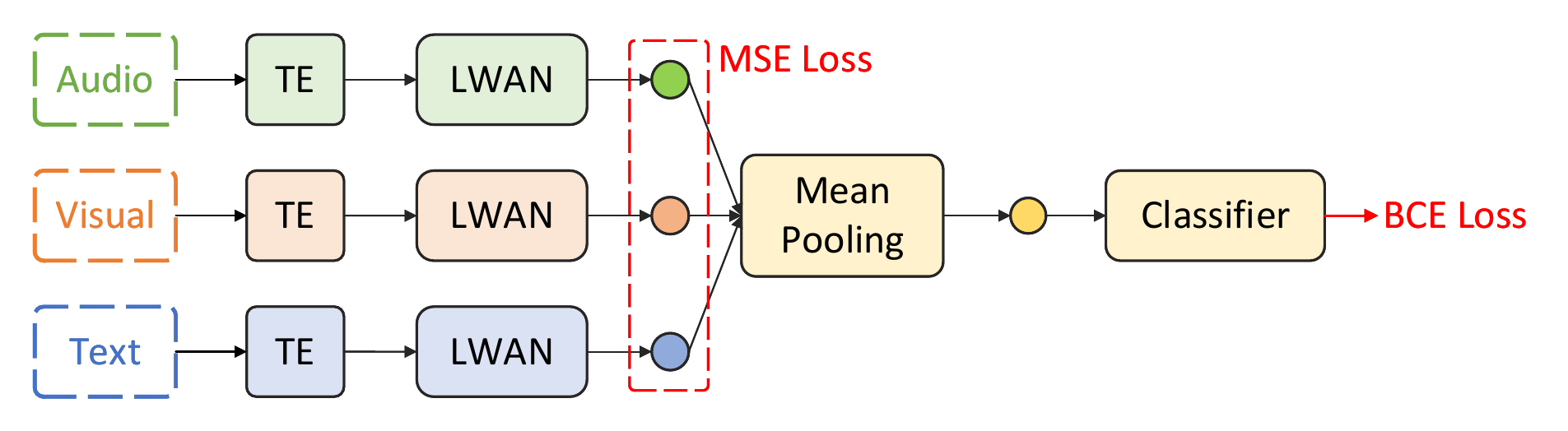}
\caption{Alignment-based fusion model}
\end{subfigure}

\begin{subfigure}{0.99\linewidth}
    \centering
    \includegraphics[width=0.9\linewidth]{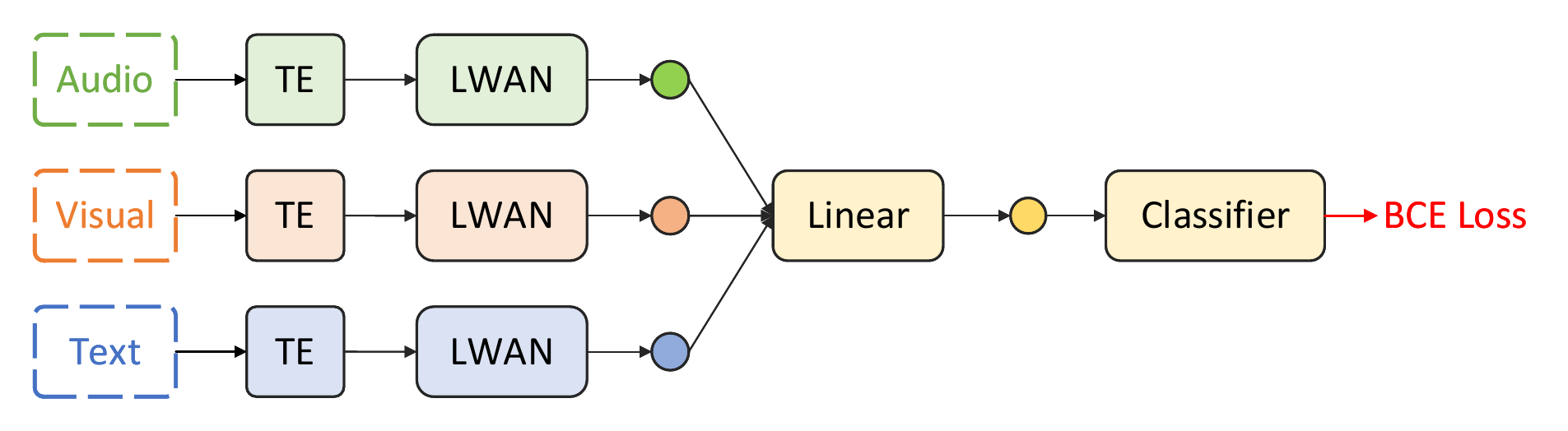}
\caption{Aggregation-based fusion model}
\end{subfigure}

\begin{subfigure}{0.99\linewidth}
    \centering
    \includegraphics[width=0.9\linewidth]{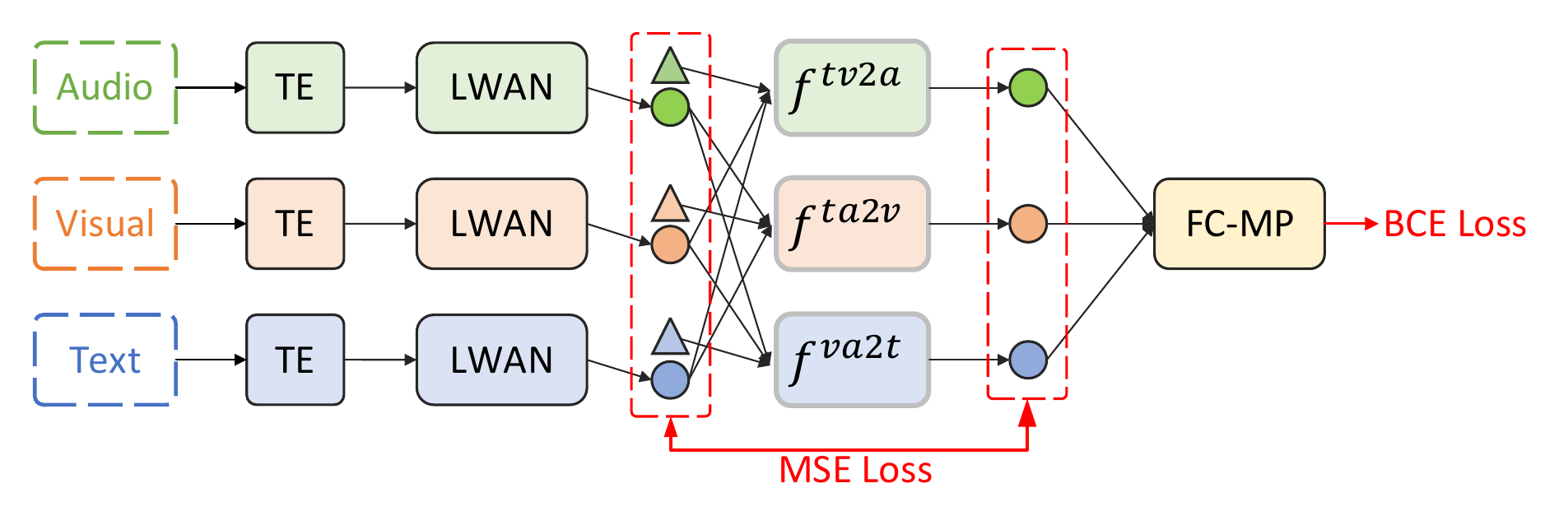}
\caption{Reconstruction-based fusion model}
\end{subfigure}
\caption{
Model structure of three different fusion mechanisms. TE denotes the Transformer Encoder, and LWAN denotes the Label-Wise Attention Network.
}
\label{fig:fusion_model}
\end{figure}

As shown in Figure \ref{fig:fusion_model}, 
we design different models for three fusion mechanisms, which simplify the CARAT framework by removing the rest except the uni-modal feature extraction and multi-modal fusion components, so as to better highlight the differences of multiple fusion mechanisms.
Specifically, the difference between the three models is mainly reflected in the fusion process after the uni-modal label-specific features $\bm{U}^{m}$ are extracted.
(1) Alignment-based fusion model, which uses MSE loss to align features in different modalities and employs Mean-Pooling to obtain a unified representation across all modalities.
(2) Aggregation-based fusion model, which utilizes a fully connected layer to directly aggregate the features of three modalities into a holistic representation.
(3) Reconstruction-based fusion model, which adopts the simplified multi-modal feature reconstruction in the CARAT framework,
specifically, it removes contrastive learning in the latent space, only retains one level reconstructing process, and uses the averaged $\bm{U}^{m}$ as the intrinsic vectors $\bm{D}^{m}$.

\begin{figure*}[h]

\centering
\includegraphics[width=0.99 \linewidth]{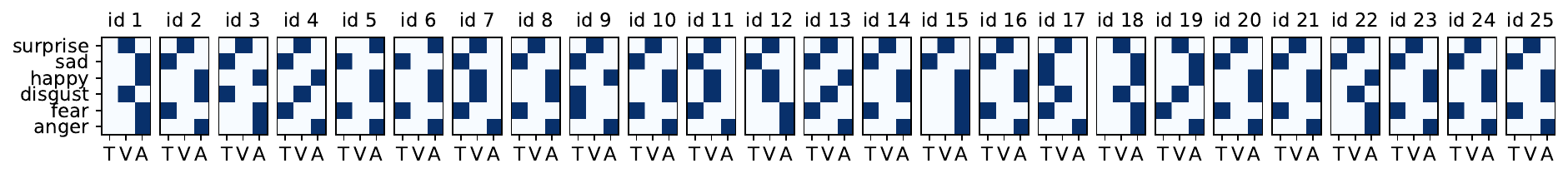} 

\includegraphics[width=0.99 \linewidth]{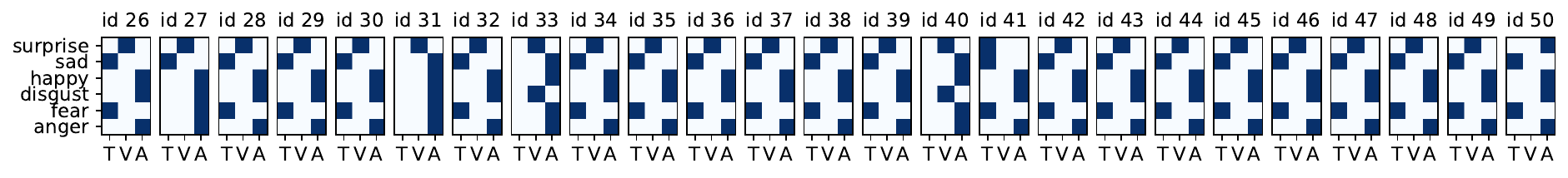} 
\caption{
The visualization of modality-to-label correlations indicates the attention of labels in each row to modality in each column across different examples.
And darker colors indicate stronger correlations. 
}
\label{fig:exp_m2l_50}
\end{figure*}

\section{E. Visualization of Modality-to-label Correlation}
\label{app:correlation}

To better reveal the correlation between labels and modalities, we present the correlation visualization of 50 random test samples, which roughly conform to the correlation rule in Figure \ref{fig:exp_m2l_freq} (a).
As shown in Figure \ref{fig:exp_m2l_50}, we can observe that:
1) Within each sample, different labels pay different attention to each modality.
2) The most relevant modality for each label will also vary from sample to sample.
3) According to all samples, each label has its own preferred modality, for example: \emph{surprise} is highly correlated with the visual modality, and \emph{sad} tends to rely on the textual modality.
All these modality-to-label correlations fit our expectations.

\end{document}